\documentclass{aa}  
\usepackage{amsmath}
\usepackage{txfonts}
\usepackage{amssymb}
\usepackage{esint}
\usepackage{natbib}
\usepackage{graphicx}
\usepackage{bm}
\usepackage[percent]{overpic}
\usepackage{multirow}

\newcommand{\dif}{\mathrm{d}}

\begin{document}

\title{Positive YORP effect induced by lateral heat conduction in a crater}
\author{Zehua Qi\inst{1,2}
	\and
	Yining Zhang\inst{1,2}
	\and 
	Hailiang Li \inst{3}
	\and
	Yangbo Xu \inst{4}
	\and
	Li-Yong Zhou\inst{1}\fnmsep\inst{2}
}
\authorrunning{Qi et al.}
\offprints{L.-Y. Zhou, \email zhouly@nju.edu.cn}
\institute{School of Astronomy and Space Science, Nanjing University, 163 Xianlin Avenue, Nanjing 210046, China
	\and
	Key Laboratory of Modern Astronomy and Astrophysics in Ministry of Education, Nanjing University, China
	\and 
    State Key Laboratory of Lunar and Planetary Sciences, Macau University of Science and Technology, Macau 999078, China
	\and 
	Shanghai Aerospace Control Technology Institute \& Shanghai Key Laboratory of Space Intelligent Control Technology, 1555 Zhongchun Road, Shanghai 201109, China
}
\date{}

\abstract{
The YORP effect plays an important role in the spin evolution of asteroids. Although craters are ubiquitous surface features, their influence on YORP torque has received limited attention. In this paper, we investigate the YORP torque of a circular crater on a spherical asteroid, focusing specifically on how lateral thermal conduction breaks symmetry to produce a net torque. Using three-dimensional finite element simulations, we calculate the resulting spin and obliquity accelerations and examine their dependence on the crater's location, depth, and thermal parameters. Our results show that the crater-induced spin torque is consistently positive, and craters at different latitudes drive the spin axis toward obliquity equilibria at $0^\circ$, $90^\circ$ or $180^\circ$. We demonstrate that the spin torque arises primarily from the lateral heat conduction inside the asteroid that occurs only in 3D model in the absence of orientation, while the contributions from self-heating and shadowing effects are negligible. While the YORP effect induced by internal heat conduction may be overtaken by torque components arising from shadowing and crater orientation—particularly on large asteroids—our numerical results show that for small craters, this spin torque amounts to approximately $\sim$10\% -- 100\% of the normal YORP torque. Its persistent positivity may help explain the observed prevalence of positive spin accelerations in asteroids. 
}

\keywords{celestial mechanics -- Asteroids -- methods: miscellaneous
}
\maketitle

\section{Introduction} \label{sec:introduction}

The scattering and re-emission of sunlight from the surface of an asymmetrically shaped asteroid may create a thermal torque, changing the asteroid's rotation period and spin axis direction in the long term. This phenomenon, known as the Yarkovsky-O’Keefe-Radzievskii-Paddack (YORP) effect \citep{Rubincam2000}, has been adopted to explain some specific distributions of spin rates and axis orientations of asteroids \citep[e.g.][]{Pravec2008, Slivan2023}.  
The YORP effect has already been directly confirmed by astronomical measurements on at least 12 asteroids \citep[e.g.][]{Taylor2007,Kaasalainen2007,vdurech2008new,Lowry2014,Hergenrother2019,vdurech2022rotation,Durech2023}. However, contrary to the expectation that the spin rate of an asteroid may speed up or slow down by a 50-50 chance, all the 12 samples are found to have positive acceleration of angular velocity. Only recently, the asteroid (433) Eros is reported to have a deceleration in rotation \citep{Feng2025}.  In fact, the numerical simulations also show that the YORP torque could increase or decrease the spin rate \citep[e.g.][]{vcapek2004yorp}, implying that the complicated process involved in the YORP effect may still not be fully understood.

Numerical simulations are often used to calculate the evolution of spin state of asteroids driven by the YORP torque. Over past decades, various numerical models have been developed based on different heat transfer models. 
With the assumption of zero thermal conductivity\footnote{Also known as Rubincam's approximation now.}, \citet{Rubincam2000} first modelled the YORP torque numerically.
Afterwards, using a one-dimensional (1D) heat transfer model, in which the heat conduction occurs only in the depth direction, \citet{vcapek2004yorp} found that the spin acceleration due to the YORP effect is nearly independent of the thermal inertia. This result was later confirmed by \citet{Breiter2008} and  \citet{Nesvorn2008} using analytical models.  
Therefore, when the global shape of an asteroid is known, the 1D model is fairly good in estimating the YORP effect, and can make accurate predictions \citep[see e.g.][for the case of (1862) Apollo]{vdurech2008new}. 

Nevertheless, \citet{Statler2009} and \citet{Breiter2009} suggested that the small-scale topography may significantly influence the YORP torque through shadowing effect. \citet{Rozitis2012} proposed the Advanced Thermophysical Model, taking the radiation blocking and re-absorption between surface elements into account, and they found that the YORP torque depends sensitively on small-scale structures on the surface even with the scale comparable to skin depth (typically a few centimetres for regolith). 

In fact, at metre scales or smaller, the influence of asteroid surface topography becomes increasingly complex. Since numerical thermophysical model requires a polyhedral shape model, the 1D assumption is valid only when the skin depth is negligible compared with the surface facet size. \citet{Golubov2012} illustrated that decimetre-sized features (such as boulders), which have been neglected in previous studies, might produce a recoil force in the tangential direction, named as tangential YORP effect (TYORP). Since thermal waves can penetrate boulders, the 1D model is no longer valid at this scale. \citet{vsevevcek2015thermal} provide an explanation of the anomalous acceleration of (25143) Itokawa through estimating the torque generated by boulders using a three-dimensional (3D) heat transfer model. More recently, \citet{Nakano2023} prove that the lateral heat conduction could dampen or enhance the YORP torque depending on different shape models.

Impact craters are common small-scale features on asteroids, and their concave geometry naturally leads to complex radiative conditions. Recently, \citet{Zhou2022, Zhou2024} found that even symmetric craters can generate a non-zero crater-induced YORP (CYORP) torque. This torque arises because the misalignment between the local normal and the radial direction of the crater, combined with shadowing and self-heating effects, breaks the symmetry of thermal re-emission from the crater surface, resulting in a net YORP torque.

Besides the geometry-driven component investigated by \cite{Zhou2022,Zhou2024}, the YORP torque of a crater may also be attributed to internal heat conduction. The latter has received less attention but may become non-negligible when the skin depth is not negligible compared with the size of craters. In this paper, we focus on metre-sized craters and investigate how lateral heat conduction contributes to crater-induced YORP effect. 
To this end, we simulate the YORP effect created by craters with a complete thermophysical model, taking into account shadowing, self-heating of thermal radiation, and lateral heat conduction. 

The rest of this paper is organized as follows. In Section\,\ref{sec:theory}, we review the basic thermal model and introduce the numerical setting of our models. In Section\,\ref{sec:result}, we investigate the dependence of YORP torque on craters' location, shape, thermal properties, and rotation period. The influence of lateral heat conduction is then explored in Section\,\ref{sect:lateralconduct}. Finally, we summarise our conclusions in Section\,\ref{sec:conclusion}. 

\section{Model and Method} \label{sec:theory}
In this paper, we use a toy model to investigate the YORP effect generated by a crater on a spherical asteroid. The YORP torque will be calculated numerically after the temperature distribution on the surface is known, while the temperature will be computed through a commercial numerical simulation software, COMSOL Multiphysics$^{\circledR}$\footnote{COMSOL Multiphysics $^{\circledR}$ v.5.6. \url{www.comsol.com}. COMSOL AB, Stockholm, Sweden.}. 

\subsection{YORP torque}
In numerical models, the asteroid surface is generally treated as a polyhedron composed of triangular facets. For an illuminated facet $i$ with area $\dif S_i$, photons re-emitting from it will impart a recoil force $\dif \mathbf{f}_i$. Assuming a Lambert surface, the recoil force of the thermal radiation is given by:
\begin{equation} \label{eq:force}
    \dif \mathbf{f}_i = -\frac{2 \epsilon \sigma T_i^4}{3c} \dif S_i\cdot \mathbf{n}_i, 
\end{equation}
where $\epsilon$ is the thermal emissivity, $\sigma$ is the Stefan-Boltzmann constant, $T_i$ is the surface temperature, $c$ is the speed of light and $\mathbf{n}$ is the unit vector normal to the surface. The net torque $\mathbf{T}$ acting on the asteroid is the summation of all the facets:
\begin{equation}
    \mathbf{T} = \sum^N \mathbf{r}_i \times \dif \mathbf{f}_i, 
\end{equation}
where $\mathbf{r}_i$ is the position vector of facet $i$ and $N$ is the total number of facets. We adopt a widely-accepted assumption that the body rotates about its principal axis of maximum moment of inertia. Thus the torque vector can be decomposed into spin component $T_{z}$ and obliquity component $T_{\varepsilon}$ and they introduce variations to the angular velocity $\omega$ and the obliquity $\varepsilon$ as \citep[see e.g.][]{vcapek2004yorp}:
\begin{equation} \label{eq:Tz}
    \frac{\dif \omega}{\dif t} = \frac{T_z}{I_z} = \frac{\mathbf{T}\cdot \mathbf{e}}{I_z}, 
\end{equation}
and
\begin{equation} \label{eq:Te}
    \frac{\dif \varepsilon}{\dif t} = \frac{T_\epsilon}{I_z\omega} = 
    \frac{\mathbf{T}\cdot \mathbf{e_1}}{I_z\omega}, 
\end{equation}
where $I_z$ is the moment of inertia around the spin axis, and the vector $\mathbf{e_1}$ is given by:
\begin{equation}
    \mathbf{e_1} = \frac{(\mathbf{N}\cdot \mathbf{e})\mathbf{e} - \mathbf{N}}{\sin\varepsilon}, 
\end{equation}
with $\mathbf{N}$ and $\mathbf{e}$ being the unit vectors normal to the orbit plane and the unit spin vector, respectively. Since the evolution time scale of YORP effect is typically much longer than the asteroid's rotation and orbital periods, the torque in Eq.\,\eqref{eq:Tz} and Eq.\,\eqref{eq:Te} can be averaged over the rotation and orbital periods.

\subsection{Basic model} \label{subsec:basicmod}

For an asteroid spinning around the principal axis, the temperature $T_i$ of facets on its surface in Eq.\,\eqref{eq:force} depends on the periodic external input of the Solar radiation and the heat conduction process inside the body. We use COMSOL$^{\circledR}$ to simulate the heat transfer and thermal radiation in fully 3D model. After obtaining the surface temperature distribution, the recoil force of thermal radiation is then calculated. The radiation interactions between surface facets including shadowing, scattering and re-absorption, as well as the lateral heat conduction beneath the surface are considered in the simulations. Distinct from the 1D model often used in previous studies, a 3D model can accurately describe the heat transfer process inside the body, particularly for small-scale structures discussed in this paper. 

We adopt a simplified model of a perfectly spherical asteroid of radius $R=10$\,m decorated with a single crater on its surface. Due to its perfect symmetry, a spherical surface produces no torque at all, thus the total torque applied on the asteroid is just the YORP torque generated by the crater. In addition, the alignment between the local normal of the spherical surface and the crater's radial vector ensures that the crater's orientation contributes no spin torque. This allows the effect of heat conduction to be studied independently of geometric contributions. 
The crater is assumed to be circular with a paraboloid-shaped profile, for which the depth $z$ as a function of radius $r$ from its centre is given by \citep{Statler2009}:
\begin{equation} \label{eq:cratershape}
    z(r) = qD \left( 1-\frac{4r^2}{D^2} \right) 
           \times 
           \begin{cases}
               1, & \text{if } r \leq D/2; \\
               \exp \left[ -\frac{(r-D/2)^2}{2\delta^2} \right], & \text{if } D/2 < r \leq D/2 + 3\delta, 
           \end{cases}
\end{equation}
where $D$ and $\delta$ are the diameter and the rim width, respectively. The maximum depth is attained at the crater centre $z_{\max}=z(r=0)$, and we define a dimensionless parameter  $q=z_{\max}/D$ to represent the depth of crater in the rest of this paper unless specified otherwise. 

Under the conditions of periodic incident radiation, the heat conduction and temperature variation mainly occur at a very shallow depth beneath the surface, while the temperature tends to be constant deep inside \citep{wesselink1948heat}. Therefore, a crater-shaped shell with a certain thickness is adopted as the thermal model of the crater. The shell is set to be isothermal at the beginning of each simulation, with an initial temperature of $280$\,K determined after some test runs. The thermal conductivity $\kappa=0.001$\,W\,m$^{-1}$\,K$^{-1}$, heat capacity $C=680$\,J\,kg$^{-1}$\,K$^{-1}$,  and surface density $\rho=1500$\,kg\,m$^{-3}$ are assumed to be constant in our simulations. These are the typical values of regolith asteroid \citep{Farinella1998}. With a diameter $D=2$\,m and a depth $q=0.2$, the crater is relatively small in size compared to the spherical asteroid of 10\,m radius. In fact, a simple algebraic calculation gives a ratio of the crater volume to the asteroid volume $\sim$$3q(D/R)^3/32$, which is $1.5\times 10^{-4}$ in this case. 

The asteroid is supposed to revolve around the Sun on a circular orbit at a heliocentric distance 1\,AU, as it's spinning around the principal axis meanwhile with a rotation period of $P=1000$\,s. We note that this rotation period is fairly short for asteroids of 10\,m radius. 

In fact, for an asteroid with given thermal parameters $(\rho, C, \kappa)$, its angular velocity $\omega =2\pi/P$ determines a parameter $\Theta$ that  is a measure of the relaxation between the absorption and re-radiation of energy \citep[e.g.][]{lagerros1996thermal1, Farinella1998,Xu2020}:
\begin{equation} \label{eq:thermalpara}
    \Theta = \frac{\sqrt{\kappa\rho C\omega}}{\epsilon\sigma T^3_{\text ss}}, 
\end{equation}
where $T_{\text ss}$ is the sub-Solar temperature. For a fast rotator with a large $\omega$, the relatively large $\Theta$ indicates a long relaxation time, thus resulting in a small temperature variation on the surface and generally a weak YORP torque. On the contrary, a high temperature gradient on the surface will be generated when $\Theta$ is low, and in this case, a much higher resolution of meshes is needed to maintain the accuracy of numerical simulations, which requires a much higher computational costs.  
Therefore, a short rotation period is adopted here, and it will not affect the qualitative results as we will show in Section\,\ref{sec:result}. This compromise between the model and computational source has been applied in previous studies \citep[e.g.][]{Xu2022, Zhang2024}.

\subsection{Mesh model} \label{subsec:meshmodel}
The finite element method used in COMSOL requires a high-quality mesh model to ensure the accuracy of simulation results. To find an appropriate mesh model, we run some test computations using a spherical asteroid model (without crater). As the asteroid rotates under the sunshine, the heat transfer happens inside the body and temperatures at different locations vary with time. Generally, the heat propagation distance can be characterised by the so-called `skin depth' or `penetration depth' 
\begin{equation} \label{skindepth}
    l_s = \sqrt{\frac{\kappa}{\omega\rho C}},
\end{equation}
which represents the depth where the amplitude of temperature variation decreases by a factor of $e^{-1}$ from the heat source (asteroid surface here). 

We assume the asteroid has the same size and other parameters as the one introduced previously in Section\,\ref{subsec:basicmod}, only except that an extra value of thermal conductivity $\kappa= 0.01$\,W\,m$^{-1}$\,K$^{-1}$ is also tested in addition to $\kappa = 0.001$\,W\,m$^{-1}$\,K$^{-1}$. 

\begin{figure}[ht]
    \centering
    \includegraphics[width=0.95\linewidth]{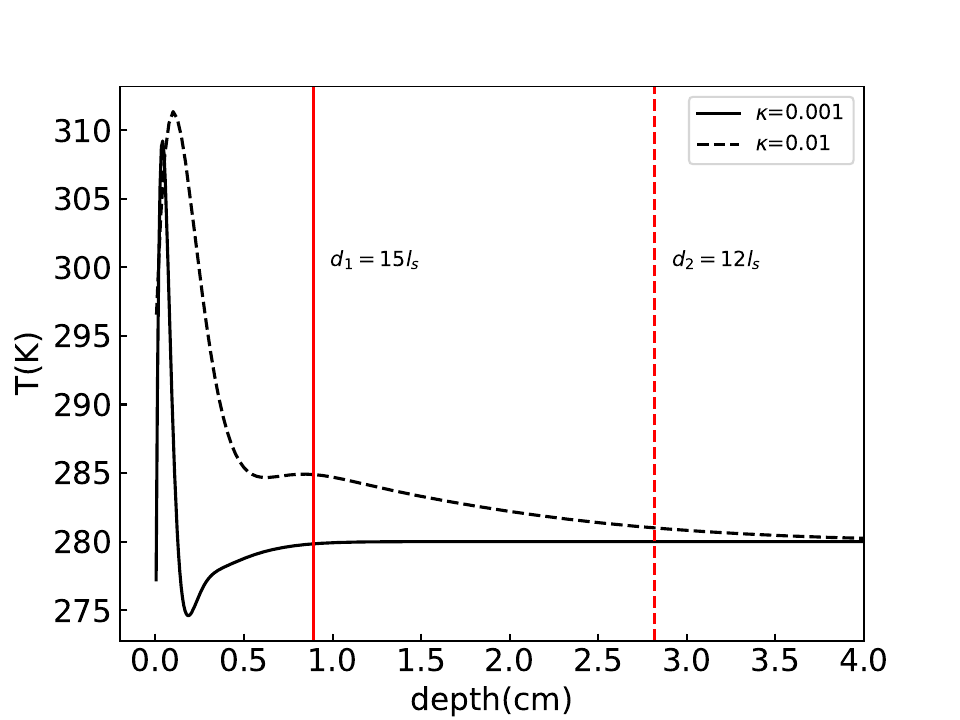}
    \caption{Temperature variations along a radius on the equatorial plane at sunset.  Two thermal conductivities $\kappa=0.001$ and $0.01$ (given in W\,m$^{-1}$\,K$^{-1}$) are adopted.  The depth is measured downward from the surface. The solid and dashed lines denote the depths, from which the temperature variation with respect to the next layer of mesh is less than $0.1\%$ for $\kappa=0.001$ and 0.01, respectively.  }
    \label{fig:skindepth}
\end{figure}

Using a mesh of triangular-prisms with a layer thickness of $l_s/2$, we simulate the temperature evolution in the asteroid by COMSOL. We select arbitrarily a point from the equator, and show in Fig.\,\ref{fig:skindepth} the temperature variation along the radius passing through this point at the moment of its sunset.  

As shown in Fig.\,\ref{fig:skindepth}, at sunset, the surface has begun to cool down, with the temperature reaching a maximum at a certain depth below the surface. As the temperature varies with depth, the temperature difference between adjacent layers of mesh decreases quickly along the depth direction. The depths, $d_1=15l_s$ and $d_2=12l_s$ for the cases of $\kappa = 0.001$ and 0.01\,W\,m$^{-1}$\,K$^{-1}$, from which the temperature variation with respect to the next deeper layer is less than 0.1\%, are indicated by solid and dashed lines in Fig.\,\ref{fig:skindepth}. Deeper than $d_1$ or $d_2$ inside the body, the temperature remains nearly constant. Therefore, to simulate the heat transfer and temperature distribution in the asteroid, we use two types of meshes in our numerical model: a high-resolution mesh in the region from the surface down to a depth of $20l_s$ ($>d_1, d_2$) to obtain accurate surface temperature, and a low-resolution mesh in the inner part to save computational cost.  

\begin{figure}
    \centering
    \includegraphics[width=0.75\linewidth]{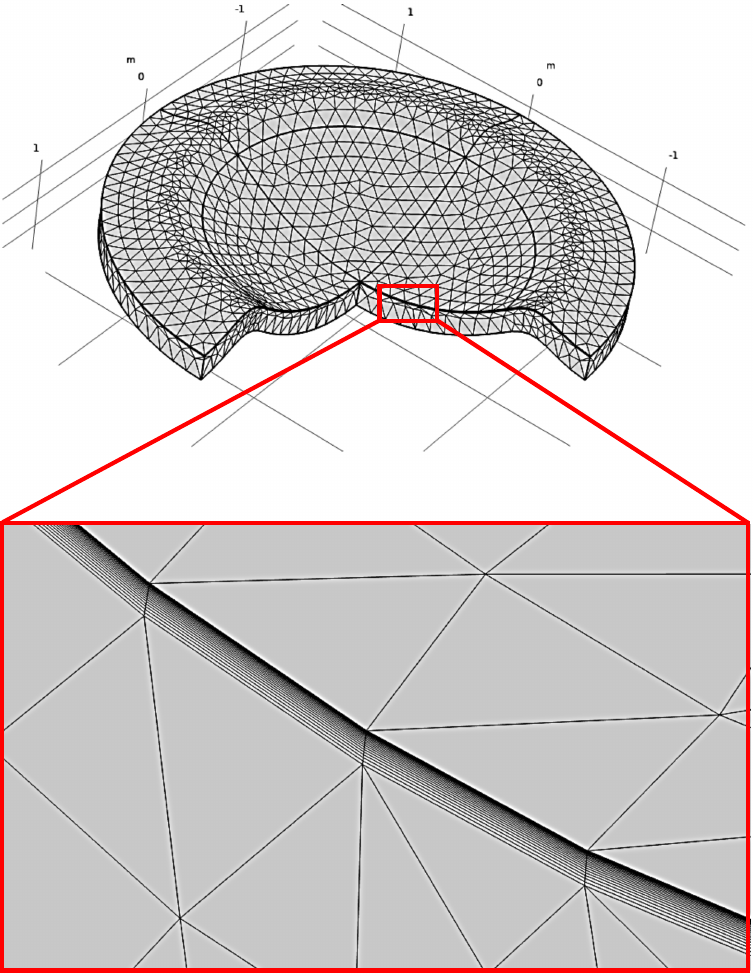}
    \caption{Mesh model for the crater. The interior is composed of tetrahedrons, wrapped by a layer of triangular prisms near the surface. }
    \label{fig:mesh}
\end{figure}

Specifically, the crater model adopted in this paper is constructed by a set of low-resolution tetrahedrons inside and a high-resolution `shell' composed of triangular prisms beneath the surface, as sketched in Fig.\,\ref{fig:mesh}. As we have shown in Fig.\,\ref{fig:skindepth}, deep inside the body, the temperature is constant, so we assume an isothermal core from a depth $100l_s$. The bottom and side edges of the crater are set to be thermally adiabatic. 

Furthermore, we regard the heat transfer equilibrium has been reached when the temperature of any surface point differs less than $0.1\%$ when it passes twice the subsolar positions in a rotation period. In our simulation, different from the widely used 1D model, the regions that lack illumination need longer time to reach a steady temperature variation pattern in the presence of lateral heat conduction. After some test runs, we find it takes 5 to 20 rotation periods to reach the equilibrium, depending on different thermal parameters. 

\section{Spin and obliquity torques} \label{sec:result}
In the simple model of a spherical asteroid decorated with a crater, the YORP effect arises only from the torque produced by the crater. In addition, compared to the asteroid, the volume and mass of the crater is very small (the ratio between them is $\sim$10$^{-4}$), therefore we simply assume that the changes in the moment of inertia and in the principal axis caused by the crater are both ignorable.

Using COMSOL, we simulate the thermal evolution in the asteroid with a crater from an initial isothermal condition, after tens of rotation periods when a dynamic equilibrium is attained inside the body, we calculate the recoil force following Eq.\,\eqref{eq:force}. Knowing the recoil force, the torque can be computed, and then averaged over the rotation and revolution periods. The angular velocity $\omega$ and the obliquity $\varepsilon$ of the asteroid might be changed by the torque, and the rates of change ($\dif \omega/\dif t$, $\dif \varepsilon/\dif t$) can be used to measure the strength of the YORP effect.

\subsection{Crater at different locations} \label{subsect:latitude}

On an asteroid spinning around an axis that is tilted by $\varepsilon$ with respect to the normal of the orbital plane, the illumination conditions of a crater vary with its location on the asteroid surface, and the YORP torque generated by it will also vary accordingly. Since the net effective torque is averaged over the rotation period, only the latitude (but not the longitude) of the crater matters. 

\begin{figure*} [htbp]
    \centering
    \resizebox{\hsize}{!}{ \includegraphics[width=1.0\linewidth]{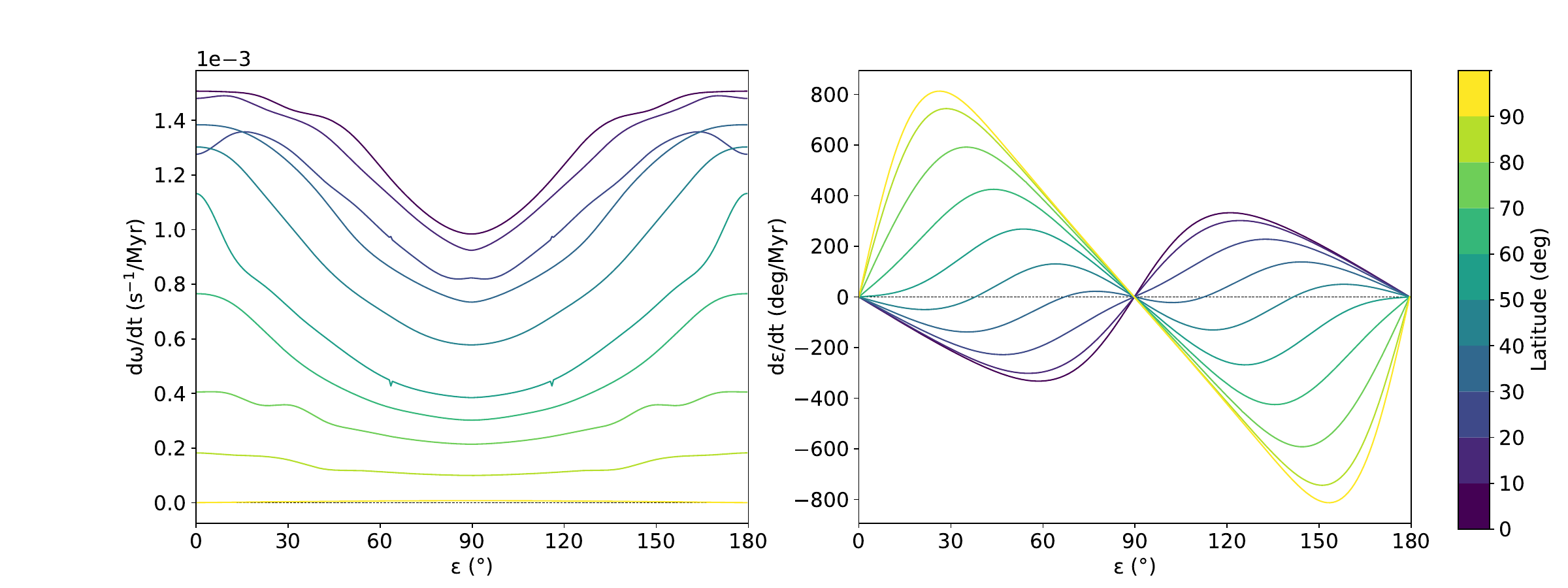}}
    \caption{YORP accelerations ($\dif \omega/\dif t$ in left panel and $\dif \varepsilon/\dif t$ in right panel) of the asteroid due to the crater as a function of obliquity $\varepsilon$. The cases of crater at ten different latitudes from $\phi=0^\circ$ to $\phi=90^\circ$ are plotted in different colours, as indicated by the colour bar. }
    \label{fig:latitude}
\end{figure*}

Setting a crater of $D=2$\,m, $q=0.2$ on different latitudes on the surface of an asteroid of $R=10$\,m, we simulate the thermal dynamics and obtain the YORP torque, from which $\dif \omega/\dif t$ and $\dif \varepsilon/\dif t$ are calculated.  We summarise the results in Fig.\,\ref{fig:latitude}.

As shown in the left panel of Fig.\,\ref{fig:latitude}, $\dif \omega/\dif t \geq 0$ for all latitudes $\phi$, indicating that the crater always accelerates the asteroid's spin rate wherever it locates. From $\phi=90^\circ$ with the crater being right at the polar and $\dif \omega/\dif t=0$, the YORP torque increases monotonically until $\phi=0^\circ$ when the crater is right on the equator. Actually, since the spin component of the torque is determined by the incident energy \citep[e.g.][]{vcapek2004yorp,Nesvorn2008,Golubov2016}, the spinning acceleration must become weaker when the crater is at higher latitude where it receives less Solar radiation because of the lower incident angle and stronger self-shadowing effect. We note that the $\dif \omega/\dif t$ reaches its maximum value at $\phi=0^\circ$ in fact also implies that the crater on the south hemisphere will accelerate the spin rate too.

For a crater at a given latitude, the spinning acceleration decreases as the obliquity $\varepsilon$ increases, and the minimum is reached at $\varepsilon=90^\circ$ when the asteroid is `lying' on the orbital plane, beyond which the variation reverses the direction for $90^\circ \leq \varepsilon \leq 180^\circ$. We note that some unexpected variations appear in our calculations, particularly for the case of $\phi=20^\circ$ and around $\varepsilon=0^\circ$ ($180^\circ$), which we believe is due to the calculation errors. 

The right panel of Fig.\,\ref{fig:latitude} shows the crater-induced variation of obliquity. In contrast to the reflectional symmetry about $\varepsilon=90^\circ$ observed in the left panel, the curves in the right panel possess rotational symmetry with respect to $\varepsilon=90^\circ$. For a prograde spinning asteroid with given obliquity ($\varepsilon<90^\circ$), the $\dif \varepsilon/\dif t$ may increase from a negative value to a positive value as the position of the crater moves from the equator to the north pole, indicating that a crater located in the polar region generally increases the obliquity while the one near the equator decreases it. 

For the asteroid with a crater at specific location (here indicated by its latitude $\phi$), the YORP effect may drive its obliquity to an equilibrium state, where $\dif\varepsilon/\dif t=0$. As shown in the right panel of Fig.\,\ref{fig:latitude}, in our calculations, when the crater is near the equator ($\phi=0^\circ, 10^\circ, 20^\circ$), $\varepsilon=0^\circ$ is a stable equilibrium, or an `asymptotic obliquity' \citep{vokrouhlicky2002yorp}, because around this point $\dif \varepsilon/\dif t <0$ when $\varepsilon>0^\circ$. While for asteroids with the crater on high latitude ($\phi\geq 50^\circ$), $\varepsilon=0^\circ$ is an unstable equilibrium, since  $\dif \varepsilon/\dif t > 0$ when $\varepsilon>0^\circ$. 

On the contrary, $\varepsilon=90^\circ$ is an unstable equilibrium in the cases of crater being near the equator, but an asymptotic obliquity for crater locating in high-latitude region. When the crater is located at mid-latitude region (in Fig.\,\ref{fig:latitude}, $\phi=30^\circ, 40^\circ$), the $\dif \varepsilon/\dif t$ curve crosses zero at $\varepsilon\sim67^\circ$ and $\sim$38$^\circ$, respectively for $\phi=30^\circ$ and $40^\circ$. This additional equilibrium is unstable, and its appearance makes both $\varepsilon=0^\circ$ and $\varepsilon=90^\circ$ asymptotic obliquities, that is, when the crater is in the mid-latitude region, for example at $\phi=30^\circ$, the YORP toque will tilt the spinning axis toward $\varepsilon=90^\circ$ if the initial obliquity $\varepsilon_0\gtrsim 67^\circ$, or toward $\varepsilon=0^\circ$ if $\varepsilon_0\lesssim 67^\circ$. 

In summary, a crater in the low-latitude region produces a YORP toque that drives the spin axis to be perpendicular to the orbital plane ($\varepsilon=0^\circ$), a crater in the high-latitude region tilts the spin axis towards the orbital plane ($\varepsilon=90^\circ$), while an asteroid with a mid-latitude ($\sim$30$^\circ$--50$^\circ$) crater has two asymptotic obliquities, either $0^\circ$ or $90^\circ$, depending on its initial state. 

It should be noted that the spin torque induced by the single crater is weak. Specifically, for $\varepsilon=0^\circ$ and $\phi=0^\circ$, the crater right on the equator produces a spin torque of $T_z^{\varepsilon, \phi=0}\sim1.2\times 10^{-8}$\,N\,m. Torques of craters on any latitudes can be estimated by interpolating between the 9 calculated values that have been used in Fig.\,\ref{fig:latitude}. Suppose 200 of such craters are distributed randomly on the asteroid. They would cover half of the total surface area, and their combined spin torque is found to be about 100 times of $T_z^{\varepsilon, \phi=0}$, that is, $T_z^{\varepsilon=0, \text{sum}}\sim1.2\times 10^{-6}$\,N\,m. \citet{Golubov2014} proposed that the YORP torque $T_z$ can be normalized as 
\begin{equation} \label{eq:nortorque}
	\tau_z = \frac{T_z c}{\Phi r_{\rm eq}^3}, 
\end{equation}
where $c, \Phi$ and $r_{\rm eq}$ are the light speed, the Solar radiation flux, and the equivalent radius of asteroid. To distinguish the boulder-induced YORP torque in tangential direction (TYORP), \citet{Golubov2014} named the torque generated by the same terrain without boulders as normal YORP (NYORP). The dimensionless normalized spin torque of our half-crater-covered asteroid is $\tau_z\sim 2.6\times10^{-4}$. This value falls within the typical range of the NYORP for real asteroids, which is on the order of $\sim 10^{-4}$ to $10^{-3}$ \citep{Golubov2014}. Therefore, for small asteroids like in our model, the crater-induced spin torque of a crater-covered asteroid is comparable to NYORP, ranging roughly from 10\% to 100\% of typical NYORP values. It is important to note that this spin torque is consistently positive, definitely leading to an acceleration of the asteroid's rotation. 

On the contrary, such randomly distributed craters are unlikely to produce a net obliquity YORP torque because the torques from the northern and southern hemispheres will typically cancel out, resulting in little to no net effect.

\subsection{Crater of different shapes} \label{subsect:shape}

For a crater of given diameter $D$ and profile defined by Eq.\,\eqref{eq:cratershape}, its shape is characterized by the depth $q$. As $q$ varies, the crater surface morphology changes, and consequently, the incident and emitted radiations vary accordingly.  Below, we numerically check the variation of YORP torque with respect to the depth of crater.

\begin{figure*}[htbp]
    \centering
    \resizebox{\hsize}{!}{ \includegraphics[width=1.0\linewidth]{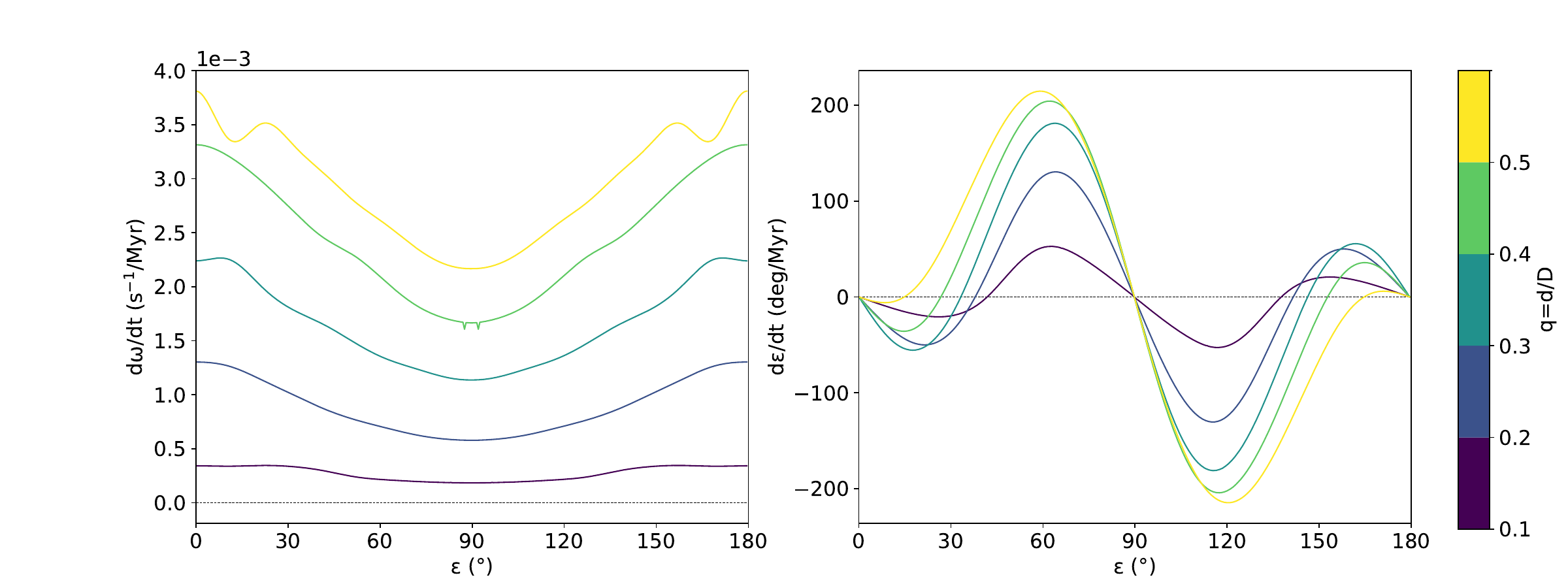}}
    \caption{Same as Fig.\,\ref{fig:latitude}, but curves are for different depths $q$ of craters as indicated by the colour bar, and the craters are located at latitude $\phi=40^\circ$.}
    \label{fig:depth}
\end{figure*}

The same model, that is, a crater of diameter $D=2$\,m on an asteroid of radius $R=10$\,m, is adopted. The location of the crater now is fixed at $\phi=40^\circ$. A crater will disappear when its depth approaches zero ($q\rightarrow 0$), while craters with $q>0.5$ (a hemispherical crater has $q=0.5$) are very unlikely to exist. Therefore, besides $q=0.2$ adopted previously in Section\,\ref{subsect:latitude}, we repeat the calculations of $\dif \omega/\dif t$ and $\dif\varepsilon/\dif t$ for other four depth values $q=0.1, 0.3, 0.4$, and $0.5$, and summarize the results in Fig.\,\ref{fig:depth}. 
 
Figure\,\ref{fig:depth} illustrates that the spin acceleration $\dif \omega/\dif t$ and the obliquity change rate $\dif\varepsilon/\dif t$ both exhibit a positive correlation with the depth $q$. Specifically, increasing $q$ from 0.1 to 0.5 results in a significant enhancement of approximately one order of magnitude for both of them. A bigger $q$ value indicates a deeper crater that has  a relatively steeper wall. And the same amount of irradiation energy absorbed and then emitted by a steep crater produces a stronger torque than a shallow crater.  

As we have shown in Fig.\,\ref{fig:latitude}, a crater on mid-latitude area gives rise to an additional unstable equilibrium of obliquity.  In the right panel of Fig.\,\ref{fig:depth}, we see a larger depth of crater drives the unstable equilibrium toward $\varepsilon=0^\circ$ (or $180^\circ$). As a result, the asymptotic obliquity of $90^\circ$ is more likely and more quickly to occur for asteroids with deep craters in this case.  

For a crater at specific location, the diameter determines the solar-illuminated and thermal-radiating areas. Therefore, the YORP torque component that arises from the overall thermal radiation along the crater's orientation, that is, the obliquity component ($\dif \varepsilon/\dif t$)  is simply proportional to the square of the diameter. But for the YORP torque ($\dif \omega/\dif t$) arising from the symmetry breaking due to internal heat conduction (see Section~\ref{sect:lateralconduct}), a smaller diameter implies that a higher degree of asymmetry may develop within the crater, although the total radiation must be correspondingly smaller. This makes the influence of the crater size on the YORP effect complicated.  

\subsection{Thermal conductivity} \label{subsect:thermal}

In a 1D heat transfer model, the heat conducts only along the depth direction, and the heat penetrates only by a few times of skin depth, $l_s$, which is generally much smaller than the size of the asteroid. For each surface element on the crater in a 1D model, it absorbs the Solar radiation, transfers the energy downwards, and then after a certain delay emits the same amount of energy through irradiation from the same surface element. Considering the symmetry of the crater,  the net spin torque ($T_z$) produced by the crater in such a 1D thermal model is in fact null. But in a 3D model, the heat diffuses in all possible directions and the spin torque leads to an acceleration $\dif \omega/\dif t$, as the numerical results (Figs.\,\ref{fig:latitude} \& \ref{fig:depth}) have shown. 

\begin{figure*}[htbp]
	\centering
	\resizebox{\hsize}{!}{ \includegraphics[width=1.0\linewidth]{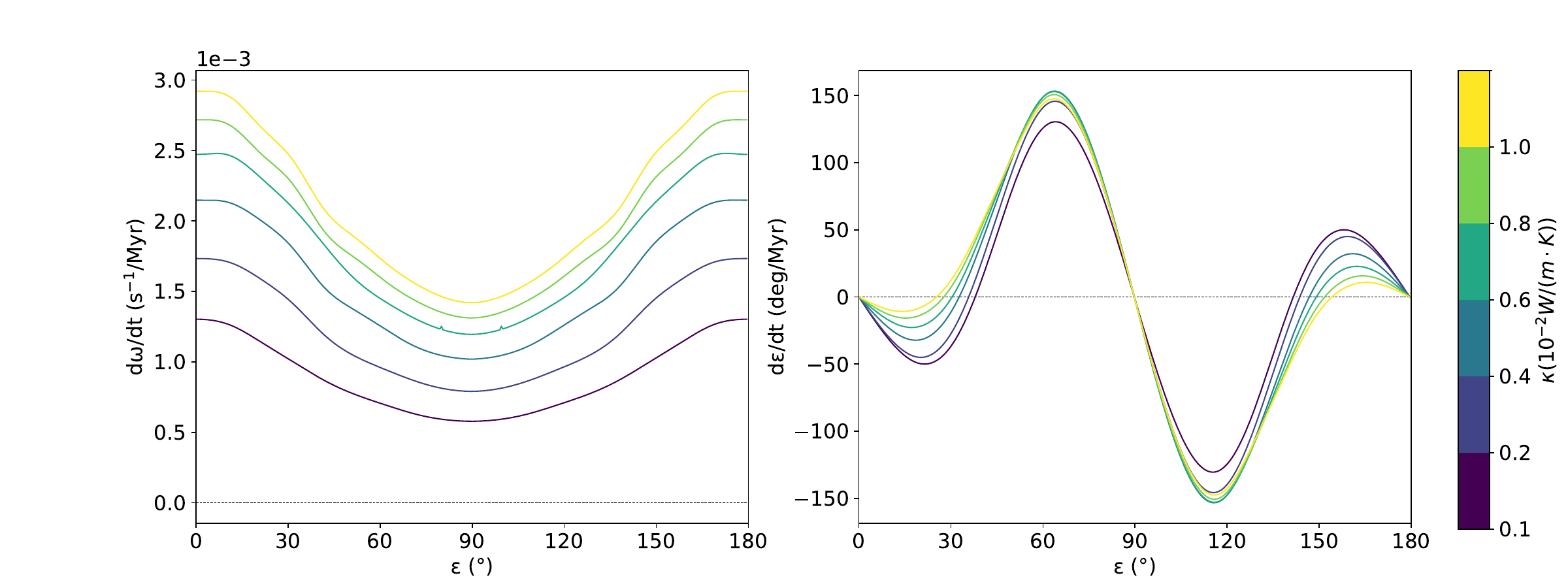}}
	\caption{Same as Fig.\,\ref{fig:latitude}. Curves of different colours are for different thermal conductivities as indicated by the colour bar. The crater has $q=0.2$ and is located at $40^\circ$. }
	\label{fig:kappa}
\end{figure*}

The thermal inertia ($\Gamma=\sqrt{\kappa\rho C}$) characterizes an asteroid's thermal response, where a high $\Gamma$ reduces temperature variations and introduces a significant `thermal lag' by delaying the thermal response to radiation. As the resulting heat transport dictates the net YORP torque, the torque produced by a crater is expected to depend on $\Gamma$. To quantify this, we perform additional calculations for a crater at $40^\circ$ latitude, varying $\Gamma$ by changing $\kappa$ between $0.001$ and $0.01$\,W\,m$^{-1}$\,K$^{-1}$ while keeping $\rho$ and $C$ fixed. The results in Fig.\,\ref{fig:kappa} illustrate this dependence.

A higher thermal conductivity facilitates faster heat diffusion in all directions, thereby creating a greater deviation from the 1D model where the net spin torque is zero. Consequently, as shown in the left panel of Fig.\,\ref{fig:kappa}, the spin acceleration $\dif \omega/\dif t$ increases with $\kappa$. Notably, the differences between adjacent curves diminish as $\kappa$ rises, suggesting that the spin acceleration approaches an upper limit. This saturation is consistent with the theoretical expectation for infinite thermal inertia ($\Gamma \rightarrow \infty$), where temperature variations vanish and the YORP torque must drop to zero. Our computations, however, were not extended to this high $\Gamma$ regime due to limited computing resources.

As shown in the right panel of Fig.\,\ref{fig:kappa}, the influence of $\kappa$ on obliquity component of the YORP effect is similar to the influence of crater depth as in Fig.\,\ref{fig:depth}, that is, a larger $\kappa$ leads to a higher probability of asymptotic obliquity at $90^\circ$. However, it is worth noting that the obliquity YORP torque exhibits considerably less sensitivity to thermal conductivity. This is evidenced by the markedly smaller variation in $\dif \varepsilon/\dif t$ with thermal conductivity in Fig.\,\ref{fig:kappa} compared to Fig.,\ref{fig:depth}. This insensitivity is further supported by the much smaller relative change in $\dif \varepsilon/\dif t$ than in $\dif \omega/\dif t$ in Fig.\,\ref{fig:kappa}.

\subsection{Rotation period}
\begin{figure}[h]
	\centering
	\includegraphics[width=0.9\linewidth]{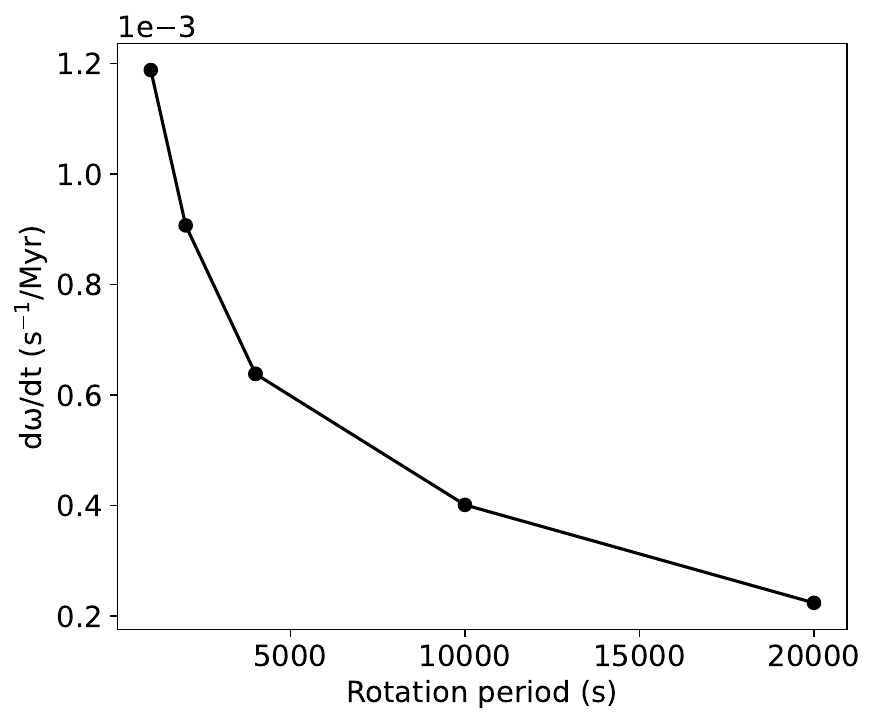}
	\caption{Spin YORP acceleration produced by a crater on equator as a function of rotation period $P$. Other model parameters are as in Fig.\,\ref{fig:latitude}.}
	\label{fig:period}
\end{figure}

So far, a short rotation period $P=1,000$\,s of the asteroid is adopted, which is a compromise between the model and computation cost. As a matter of fact, the steady-state rotation periods of rubble-pile asteroids typically fall within the range of hours. Keeping other parameters of the model as before but varying the rotation period in 2,000-20,000\,s, we calculate the spin YORP torque of a crater locating on the equator. As summarized in Fig.\,\ref{fig:period}, the spin torque decreases with rotation period. For $P= 20,000$\,s, the $\dif \omega/\dif t$ is nearly one order of magnitude smaller than that for $P=1,000$\,s.

The thermophysical model \citet{lagerros1996thermal1} suggests that the surface temperature is determined by the thermal parameter $\Theta$ once the shape is fixed, while $\Theta$ represents the ability to maintain the temperature during periodic external insolation \citep{Harris_2020}. When the rotation period is fairly long for large asteroids, their $\Theta \approx 0$, and the spin YORP effect is dominated by geometric asymmetry rather than the  internal heat conduction as in the case of this paper.

\section{Lateral heat conduction} \label{sect:lateralconduct}

For any structure on an asteroid's surface, because the lever arm and normal direction of any surface element $\dif S$ remain constant in a body-fixed frame, the rotation-averaged torque produced by $\dif S$ solely depends on the re-emitted energy, which is nearly equal to the total radiation received by the same surface element. In this scenario, the spin YORP torque is nearly independent of the thermal inertia \citep{Nesvorn2008}, and the symmetric topology of the crater in our model will not generate the spin YORP torque. 

However, the lateral heat conduction in the 3D model might break the balance of energy budget. Consequently, the numerical results demonstrate a positive spin acceleration $\dif \omega/\dif t$ (Fig.\,\ref{fig:latitude}) and its dependence on the thermal inertia (Fig.\,\ref{fig:kappa}). The lateral heat conduction inside the body plays a critical role in creating the spin torque, as we will show below.

\subsection{Asymmetry in temperature distribution and YORP torque}

For any surface element on the crater, the perfect equality between the absorbed Solar radiation and the re-emitted thermal radiation could break if either the lateral heat conduction or the self-heating occurs. Particularly, a spin YORP torque may arise from the asymmetries in both the spatial temperature distribution and its temporal variation between the eastern and western sides of the crater. To show this, we located a crater of diameter $D=2$\,m and depth parameter $q=0.2$ on the equator of an asteroid with radius 10\,m. The north and south halves of such a crater are absolutely symmetrical to each other, and there will be no obliquity YORP torque. 

\begin{figure}[ht]
	\centering
	\includegraphics[width=1.0\linewidth]{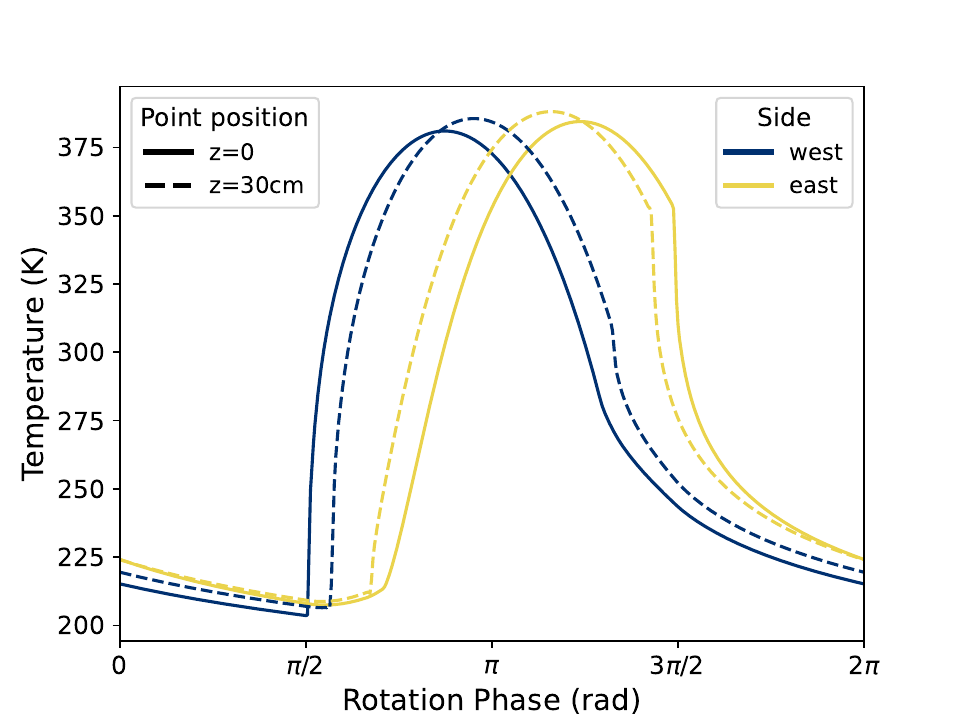}
	\caption{Temperature variations at symmetrical points on the crater's prime vertical circle. Solid and dashed lines correspond to points at depths of $z = 0$ and $z = 30\,\text{cm}$, respectively, while blue and yellow colours indicate locations on the west and east sides of the crater.}
	\label{fig:tempvar}
\end{figure}

Arbitrarily, we selected two point pairs on the crater's prime vertical circle (the great circle passing through the east, nadir and west points): one at $(r = D/2, z = 0)$ and the other at $(r = D/4, z = 3D/20)$. According to Eq.\,\eqref{eq:cratershape}, the first pair lies on the crater's edge where it meets the rim, while the second is located lower on the crater wall at a depth of $z = 30$\,cm. The temperatures at these four points were calculated in the 3D model and plotted over one rotational period in Fig.\,\ref{fig:tempvar}, with the rotational phase starting from midnight.  
 
During the local night, temperatures at all four points decline slowly under radiative cooling, then surge abruptly at morning illumination. The west-side points, receiving sunlight first at a high incidence angle, warm rapidly, triggering lateral subsurface heat conduction toward the east. As a result, the east-side points experience pre-sunrise warming due to both this heat transfer and the self-heating effect. At local sunrise, the east-side temperatures rise more gradually owing to the lower Solar incidence angle, taking longer to peak. Additionally, the shadowing effect causes the east-side points to lose direct Solar illumination at a higher incidence angle, leading to an abrupt temperature drop around the rotational phase of $3\pi/2$. As summarized in Fig.\,\ref{fig:tempvar}, this asymmetry creates not only a time lag but also a pronounced difference in the temperature profiles between the two sides.

For symmetrically placed points on the east and west sides, the west point exceeds the east in temperature for only about a quarter of the rotational period. Furthermore, the east point exhibits higher minimum and maximum temperatures than its western counterpart. For instance, as shown in Fig.\,\ref{fig:tempvar} for points at the rim ($z=0$), the west point's temperature ranges from 203.6\,K to 381.0\,K, whereas the east point's varies from 207.5\,K to 384.5\,K.

\begin{figure}[ht]
	\centering
	\includegraphics[width=1.0\linewidth]{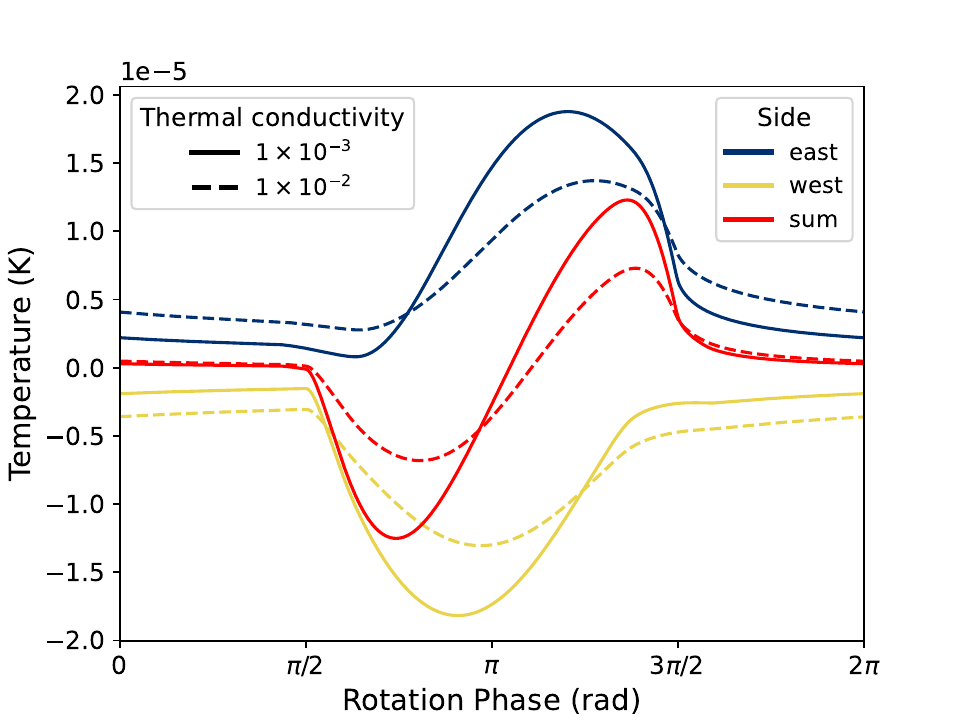}
	\caption{Spin YORP torques produced by the western (yellow) and eastern (blue) halves of the crater. The total torque (the summation of both halves) is also plotted in red. The solid and dashed lines are for thermal conductivity $\kappa=0.001$ and $0.01$\,W\,m$^{-1}$\,K$^{-1}$, respectively.}
	\label{fig:torque2sides}
\end{figure}

A higher surface temperature leads to a stronger thermal radiation recoil force. This results in an imbalance where the positive spin YORP torque generated by the eastern half may exceed the negative torque from the western half. To quantify this, we computed the spin YORP torques from the two sectors separated by the crater's meridian. Their variations over a rotational period are shown in Fig.\,\ref{fig:torque2sides}. Since this torque asymmetry originates partly from lateral heat conduction, an effect enhanced under higher thermal conductivity $\kappa$, we present calculations for two $\kappa$ values for comparison. 

Figure\,\ref{fig:torque2sides} clearly reveals a time lag between the positive spin torque (from the east half) and the negative torque (from the west half). A more subtle but crucial feature is the asymmetry in their temporal profiles. Compared to the case of $\kappa=0.001$\,W\,m$^{-1}$\,K$^{-1}$, a higher thermal conductivity $\kappa=0.01$\,W\,m$^{-1}$\,K$^{-1}$ (dashed lines) strengthens the internal heat conduction, dampening the surface temperature variation. This reduces the peak torque magnitudes for both crater halves. Furthermore, enhanced conductivity allows more heat to be stored internally and released as thermal radiation during night (approximately in phases $(0, \pi/2)$ and $(3\pi/2, 2\pi)$). This effect produces a distinctive `tail' in the net positive spin torque (red lines) after sunset (phase $3\pi/2$), which constitutes the primary contribution to the total averaged YORP torque. Of course, it is worth noting that the torque might disappear also in the extreme but highly improbable case of an extremely large thermal conductivity or a very small asteroid, where the temperature variation would be negligible.

\subsection{A pseudo-crater model} \label{subsect: pseudo-crater}

To elucidate how a positive YORP torque can arise from a symmetric structure like a crater, we introduce a `pseudo-crater' model. This model replaces the crater in Fig.\,\ref{fig:mesh} with a semi-cylindrical shell of radius $r=0.1$\,m, oriented north-south on the equator of a spherical asteroid with radius $R=10$\,m. As the asteroid rotates, every point along a cylindrical generatrix of this shell experiences identical Solar radiation conditions. Consequently, no temperature gradient is established in this longitudinal direction, and thus, no heat diffusion occurs along it. Effectively, this reduces the system to a 2D model where the heat diffusion is confined to the east-west direction. For simplicity, we assign the shell a unit length ($h=1$\,m) and the same thermal parameters as the crater model used previously in this paper. Thermal diffusion and radiation from the cylinder's top and bottom ends are neglected.

\begin{figure}[h]
    \centering
    \includegraphics[width=1.0\linewidth]{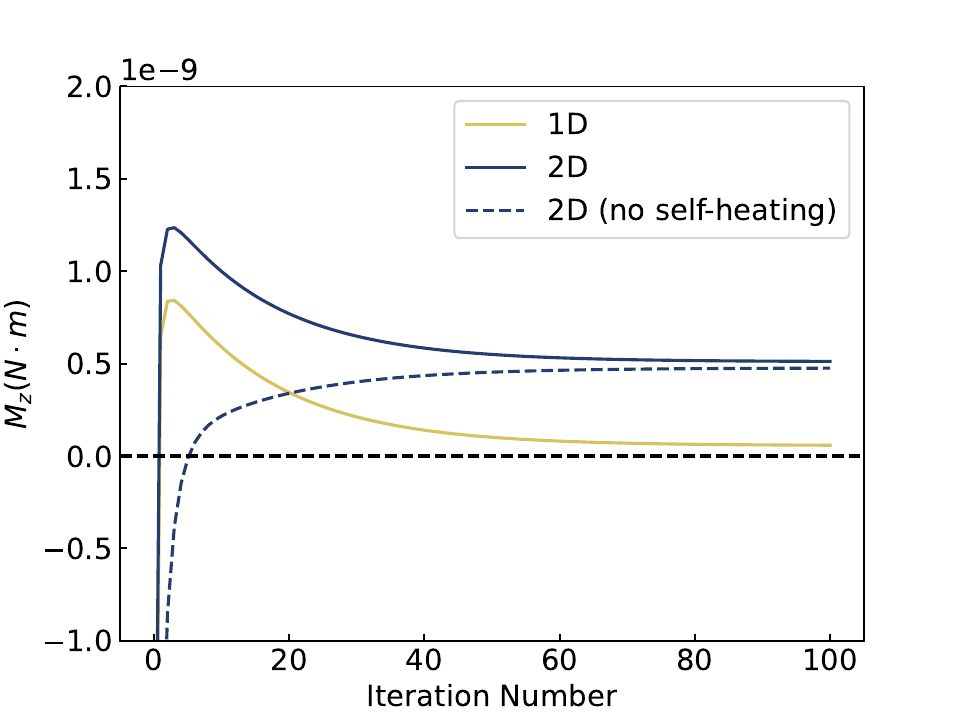}
    \caption{Convergence of calculated YORP torque produced by the pseudo-crater. The iteration number indicates the number of rotation periods of the asteroid counting from the beginning of our calculation where an isothermal state was assumed. }
    \label{fig:dimension}
\end{figure}

To evaluate the YORP torque generated by the pseudo-crater, we computed the surface temperature distribution starting from an isothermal state, and subsequently the spin YORP torque was obtained by averaging the torque of recoil force over an entire asteroid rotation. We employed three distinct thermal models: a 1D model confined to vertical heat conduction, a 2D model incorporating lateral conduction (only in east-west direction) but neglecting the self-heating, and a full 2D model accounting for both lateral conduction and self-heating. A comparison of the resulting torques (Fig.\,\ref{fig:dimension}) may help us isolate the contribution of lateral heat diffusion and self-heating.

Starting from an isothermal state, both the eastern and western facets lose energy through thermal radiation before sunrise. However, the eastern facet, which receives sunlight later, cools more and thus begins the first rotation at a lower temperature. Consequently, it stores more solar energy while emits less thermal radiation than the western facet, resulting in an initial negative spin torque in both 1D and 2D models. From the second rotation onward, self-heating pre-warms the eastern facet before sunrise. This process transfers extra energy that must be radiated away, generating a positive net spin torque. In contrast, models without self-heating lack this energy transfer; as Fig.\,\ref{fig:dimension} shows, the net torque gradually converges to a certain value as heat diffuses inside the body toward a dynamic equilibrium.

After the initial rotation, the self-heating effects on the eastern and western crater facets become largely symmetric as the temperature difference between two sides diminishes. The net torque is thereby dominated by direct Solar irradiation and internal thermal conduction. This is confirmed by the convergence of the time-averaged torques to stable values at dynamic equilibrium (Fig.\,\ref{fig:dimension}). The close agreement between the final values in the 2D model without self-heating ($4.75\times 10^{-10}$\,N\,m) and the full 2D model ($5.12\times 10^{-10}$\,N\,m) strongly suggests that self-heating is inconsequential for the overall YORP torque in this scenario.

In contrast to the 2D models, the 1D model yields a final YORP torque of only $5.84 \times 10^{-11}$\,N\,m, nearly an order of magnitude smaller and effectively negligible. A separate 1D simulation without self-heating (not shown) converges to a nearly identical value, confirming that the pseudo-crater generates only negligible spin torque when heat diffusion is restricted to the depth direction. The dramatic increase in torque observed in the 2D models unequivocally identifies lateral heat transfer as the primary driver of the spin YORP effect in this crater structure.

Since numerical experiments demonstrate that the spin YORP torque is mainly attributable to the lateral heat transfer within the body, it follows that this torque will vary with the thermal conductivity $\kappa$ in a given model. Further tests with different $\kappa$ values in the same pseudo crater model revealed a notable dependence of the spin YORP torque on this parameter. In fact, when $\kappa$ is very small, the asymmetric temperature distribution caused by thermal lag is negligible. On the other hand, the temperature difference between the west and east sides vanishes when $\kappa$ is very large. Therefore, the maximal spin YORP torque is attained at some specific $\kappa$ in between, which was found to be $\kappa\sim 2\times 10^{-2}$\,W\,m$^{-1}$\,K$^{-1}$, corresponding to a skin depth $l_s\approx 0.2$\,cm, in our preliminary calculations. We leave a detailed investigation on the dependence of this torque on thermal parameters to a separated paper.

\section{Conclusions} \label{sec:conclusion}
While the YORP effect is known to depend on an asteroid's macroscopic shape and thermal parameters, its sensitivity to small-scale surface features, including boulders, craters, and roughness, is also recognized \citep[e.g.][]{Breiter2009, Statler2009, Golubov2012, Rozitis2012}. Recent studies of the crater-induced YORP effect (CYORP) showed that the YORP torque may stem from the crater's orientational asymmetry \citep{Zhou2022,Zhou2024}. In these studies, the lateral heat conduction inside the asteroid is often ignored, thus a fully symmetrical configuration, for example a rotationally symmetrical crater located on the equator of a spherical asteroid, produces no net torque. However, when the internal heat conduction (an ever-present factor in reality) is considered, a net YORP torque can emerge. In this paper, we employed a complete 3D model to calculate the radiation interactions between surfaces and heat conduction within the body, and we demonstrated that the heat conduction can indeed generate a net YORP torque, even in perfectly symmetrical crater geometries.

To eliminate the influence of geometric characteristics, we adopted a simple model of spherical asteroid featuring a circular crater with a paraboloid profile bounded by a raised rim. The radii of the asteroid and crater are set to $R=10$\,m and $r=1$\,m, respectively, while the crater morphology is characterized by its depth-to-diameter ratio $q$. Under the assumption of zero thermal inertia, this symmetrical configuration generates a torque in obliquity but yields no net spin torque over a full rotation. 

Using typical thermal parameters, we numerically solved the 3D heat conduction equation to obtain the temperature distribution on the crater's surface. From this, the recoil force and YORP torque were calculated. Our results showed that the spin torque generated by the crater is consistently positive, with its magnitude increasing monotonically as the crater's location shifts from the polar to the equatorial region (Fig.\,\ref{fig:latitude}). The obliquity YORP torque, however, demonstrates a non-monotonic dependence on crater location. A low-latitude crater drives the obliquity toward an asymptotic equilibrium at $\varepsilon=0^\circ$. In contrast, a high-latitude crater drives it toward $\varepsilon=90^\circ$. In the mid-latitudes (for example, $\phi=30^\circ$ and $40^\circ$), both $0^\circ$ and $90^\circ$ are stable equilibrium points for obliquity $\varepsilon$ (Fig.\,\ref{fig:latitude}).

The crater's geometry also affects the torque. As the depth parameter $q$ increases from 0.1 to 0.5, the changing illumination conditions within the crater alter both the spin and obliquity torques by one order of magnitude. Specifically, a larger $q$ indicates steeper crater walls, leading to a larger temperature gradient on the surface. Both the heat conduction within the body and the shadowing effect become more pronounced at higher $q$, and consequently both the spin and obliquity torques increase (Fig.\,\ref{fig:depth}). 

In our model of spherical asteroid decorated with a symmetrical crater, the obliquity torque is primarily produced by the overall thermal radiation from the crater, which is determined mainly by the crater's orientation (aligned with the position direction) and the shadowing effect. The strength of these obliquity torques, measured in normalized torque, range from $\sim$10$^{-6}$ to $10^{-3}$, which are consistent with the findings of \citet{Zhou2022}. However, the spin torque, which is about two orders of magnitude smaller than the obliquity torque (see Fig.\,\ref{fig:latitude} and Eqs.\,\eqref{eq:Tz} \& \eqref{eq:Te}), arises mainly from the symmetry breaking due to lateral heat conduction beneath the crater's surface. Therefore, an increase in thermal conductivity enhances the spin torque significantly, while the obliquity torque is affected only marginally (Fig.\,\ref{fig:kappa}).

For the same reason, crater size (diameter) directly affects the obliquity torque—it is approximately proportional to the square of the diameter. The dependence of the spin torque on diameter, however, is somewhat more complex. For sufficiently large craters, the asymmetry in surface temperature distribution caused by lateral heat conduction approaches an equilibrium, such that the spin torque tends to an asymptotic value $M_z^{\infty}$. Our preliminary calculations indicate that the spin torque $M_z$ depends on the crater diameter $D$ as $M_z \approx M_z^{\infty} / \left[1 + (D_0 / D)^{3/2}\right]$. A detailed investigation is left for future work. 

In our model, we set a rotation period $P=1,000$\,s. For longer periods, the results remain qualitatively the same. Quantitatively, the spin YORP torque  decreases to about one-sixths as $P$ increases from 1,000\,s to 20,000\,s  (Fig.\,\ref{fig:period}).

A crater on a spherical asteroid equator, with its orientation aligned with the local surface normal, produces no obliquity torque but it serves as a clear illustration of spin torque generation. We calculated the temperatures at representative points in such a crater (Fig.\,\ref{fig:tempvar}), and the resulting spin torques from thermal radiation on its eastern and western halves (Fig.\,\ref{fig:torque2sides}). The temporal profiles of these temperatures and torques reveal how an asymmetric temperature distribution is established in such a symmetrical crater configuration, and ultimately a net positive spin torque is produced.

To isolate the essential influence of lateral heat diffusion on the generation of the spin YORP torque, we constructed a pseudo-crater model where the temperature gradient was confined to the east-west direction. Our numerical simulations demonstrated that a spin torque arises exclusively when the lateral heat conduction is allowed (in a 2D model), with the self-heating effect making only a negligible contribution (Fig.\,\ref{fig:dimension}).  

We have quantified the spin YORP torque arising from lateral heat conduction in a crater using the dimensionless normalized torque and found it to be comparable to the normal YORP. Consequently, the cumulative effect from multiple craters is non-negligible. A rough estimate suggests that for a heavily cratered asteroid, the cumulative contribution from craters could range from 10\% to 100\% of the normal YORP torque.

In our calculations, the orientation effect of the crater was not considered. We assumed that the crater orientation is always aligned with the local  surface normal. It is worth noting that the obliquity component of YORP torque produced by craters located off the equator (right panels of Figs.\,\ref{fig:latitude}, \ref{fig:depth} \& \ref{fig:kappa}) is, to some extent, physically equivalent to the torque generated by craters with different orientations, as studied in \citet{Zhou2022}. 

We note that the calculations in this paper are mainly based on an ideal model, in which a crater of radius 1\,m sits on an asteroid of radius 10\,m. The lateral-heat-conduction-induced spin YORP torque might be suppressed by geometric effects for large craters and large asteroids.  The real asteroid is often more complex due to the combined contribution of TYORP, CYORP, NYORP, and boulder-induced YORP \citep{baker2025boulder}. The general solution of geometric and thermophysical effects across scales is therefore a key direction for our future studies.

\begin{acknowledgements}
 This work has been supported by the National Natural Science Foundation of China (NSFC, Grants No.12373081 \& No.12150009) and the China Manned Space Program with grant No.CMS-CSST-2025-A16.
\end{acknowledgements}

\bibliographystyle{aa.bst}
\bibliography{reference}

\end{document}